# Voltage-controlled magnetism enabled by resistive switching

Pavel Salev[1], Iana Volvach[2], Dayne Sasaki[3], Pavel Lapa[1],
Yayoi Takamura[3], Vitaliy Lomakin[2], Ivan K Schuller[1]

[1]Department of Physics, University of California San Diego
[2]Center for Memory and Recording Research, University of California San Diego
[3]Department of Materials Science and Engineering, University of California Davis

The discovery of new mechanisms of controlling magnetic properties by electric fields or currents furthers the fundamental understanding of magnetism and has important implications for practical use. Here, we present a novel approach of utilizing resistive switching to control magnetic anisotropy. We study a ferromagnetic oxide that exhibits an electrically triggered metal-to-insulator phase transition producing a volatile resistive switching. This switching occurs in a characteristic spatial pattern: the formation of a transverse insulating barrier inside a metallic matrix resulting in an unusual ferromagnetic/paramagnetic/ferromagnetic configuration. We found that the formation of this voltage-driven paramagnetic insulating barrier is accompanied by the emergence of a strong uniaxial magnetic anisotropy that overpowers the intrinsic material anisotropy. Our results demonstrate that resistive switching is an effective tool for manipulating magnetic properties. Because resistive switching can be induced in a very broad range of materials, our findings could enable a new class of voltage-controlled magnetism systems.

Achieving efficient voltage-controlled magnetism remains one of the central efforts in the scientific and engineering communities[1–3]. Electrical manipulation of magnetic parameters such as magnetization or Néel vector direction, magnetic anisotropy, coercivity, saturation magnetization, transition temperature, exchange bias field, etc., is crucial for building spintronic devices for applications in next generation information technology[4,5]. Many approaches have been developed for electrical control of magnetism including devices based on intrinsic multiferroic materials[6–8] or artificial magneto-electric heterostructures[9–11], field-effect devices with solid[12–14] or electrolyte[15–17] gates, magneto-ionic systems[18–20] and exchange bias heterostructures[21–23]. Despite promising results, those approaches often suffer from a range of drawbacks such as high-power consumption and the need of an external magnetic field for deterministic operation. Thus an avid interest remains in discovering new ways to achieve voltage-controlled magnetism. In this work, we present a novel concept of utilizing resistive switching to enable voltage-controlled magnetic anisotropy (VCMA). Resistive switching is the ability to change the material's resistivity by application of voltage/current. This phenomenon has attracted enormous attention in the past decades because of the possible applications in binary memories for conventional computers and analog, synapse-like memories for hardware level neuromorphic computing[24,25]. Typically, resistive switching occurs in characteristic spatial patterns resulting in the coexistence of conducting/insulating regions inside the materials. A common example of such a pattern is a conducting filament that percolates through an insulating matrix[26,27]. Here we show that the resistive-switching-driven formation of conducting/insulating regions in a magnetic oxide device, which creates the concomitant pattern of ferromagnetic (FM) / paramagnetic (PM) regions, results in the development of a strong uniaxial magnetic anisotropy. This anisotropy development closely follows the switching and, thus, can be turned on and off reversibly by voltage. Our work shows that employing resistive switching is a viable strategy to achieve voltage-controlled magnetism, which potentially could be implemented in a variety of ferromagnetic, ferrimagnetic, and antiferromagnetic materials.

We explored the resistive switching impact on magnetic properties in $La_{0.7}Sr_{0.3}MnO_3$ (LSMO) using planar $50\times100$ μm² two-terminal devices. A special property of LSMO is the phase transition at $T_c \approx 340$ K from a low-temperature FM metal to a high-temperature PM insulator. Our recent work[28] showed that applying voltage to an LSMO device can trigger those coupled metal-insulator and FM transitions resulting in a volatile resistive switching, i.e. a switching that automatically resets upon removal of the applied voltage. On a microscopic level, the metal-to-insulator switching in LSMO occurs by the formation of a transverse PM insulating barrier that spans across the entire device width and impedes the electric current flow[28]. Figs. 1 a and b illustrate the barrier formation process in an LSMO device. The I-V curve of the device (Fig. 1a) is highly nonlinear and shows the development of a negative differential resistance (NDR) region above 13 V, i.e. the part of the curve where the slope *dV/dI < 0*, which is a signature of resistive switching. Spatially resolved magneto-optical Kerr effect (MOKE) measurements showed that during the resistive switching, the uniform FM state across the device (Fig. 1b, blue line) transforms into a state with a nonmagnetic barrier located in the device center (Fig. 1b, red line). Inside this barrier, the MOKE signal completely vanishes. On the left and right sides of the barrier, however, the device remains FM and the MOKE signal is comparable to that in equilibrium (i.e., at zero voltage). The evolution of magnetic anisotropy in the FM regions surrounding the PM barrier when the device undergoes resistive switching is in the focus of this paper.

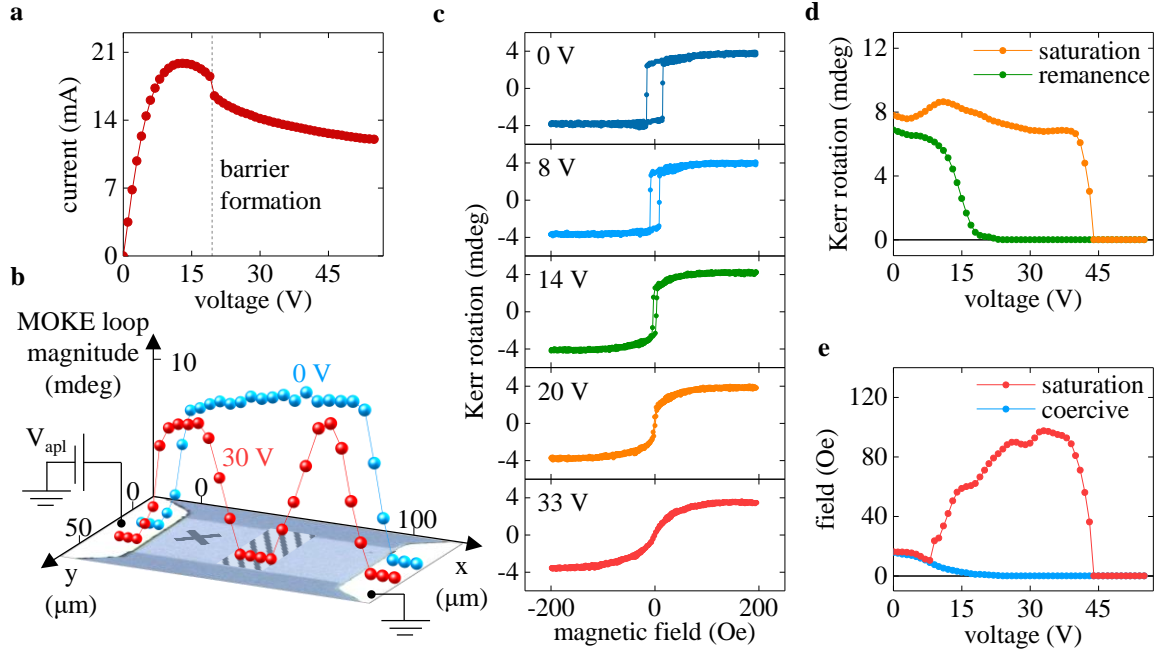

**Fig. 1 | Voltage-controlled magnetism driven by resistive switching. a**, I-V curve of a 50×100 μm$^2$ LSMO device. The development of a negative differential resistance region, *dV/dI < 0*, indicates the volatile metal-to-insulator resistive switching. **b**, MOKE hysteresis loop peak-to-peak amplitudes recorded along the device length in equilibrium (0 V, blue line) and during the resistive switching (30 V, red line). The curves are overlaid with an optical image of the device (light blue rectangle is the LSMO bridge, white areas are electrodes). Non-zero values of the Kerr signal correspond to the FM regions. The zero signal at 30 V at the device center indicates the formation of a PM insulating barrier (highlighted by a patterned rectangle). **c**, MOKE hysteresis loops recorded at a spot close to one of the electrodes (marked by a cross in **b**) at several applied voltages. As the voltage drives the device through the resistive switching, the loops become oblique indicating the change of magnetic anisotropy. **d**, **e**, MOKE signals at saturation and remanence signals (**d**) and coercive and saturation fields (**e**) as functions of the applied voltage. As the device undergoes the resistive switching, the remanence suppression and saturation field increase can be observed. All measurements were performed at 100 K.

Resistive switching in LSMO produces a strong VCMA effect as evidenced by the pronounced change of the hysteresis loop shape. Fig. 1c shows the evolution of the MOKE hysteresis loops as a function of applied voltage. The loops were recorded using a ~5-μm-size laser beam focused at a spot close to one of the electrodes (marked by a cross in Fig. 1b). The measurement spot was approximately 40 μm away from the position where the PM insulating barrier first appears during the resistive switching. The magnetic field was applied along the device length. The measurements were performed at 100 K. While ramping up the voltage in the 0-8 V range, we observed typical square-shaped hysteresis loops. The only noticeable change in this 0-8 V range is ~40% coercivity reduction, which is most likely due to the uniform Joule heating that is expected to precede the resistive switching[28]. For applied voltages above 8 V, the resistive switching initiates as indicated by the development of nonlinearities in the I-V curve (Fig. 1a). At the same time, the character of the hysteresis shape changes: the initial square loop becomes oblique stretching along the field axis. After the formation of PM insulating barrier above 20 V, the coercivity and remanence vanish resulting in a non-hysteretic MOKE loop. Because the resistive switching in our LSMO devices is volatile, removing the applied voltage resets the device restoring the initial square hysteresis loop and the whole process can be repeated. The observed change of the hysteresis loop from square to oblique shape indicates the development of a hard magnetic anisotropy along the device length during the voltage-driven resistive switching. The overall impact of this voltage-induced anisotropy in the LSMO devices, i.e., the complete suppression of coercivity and remanence, can be qualitatively compared to the best examples of VCMA in magneto-electric[10] and magneto-ionic[18,19] devices.

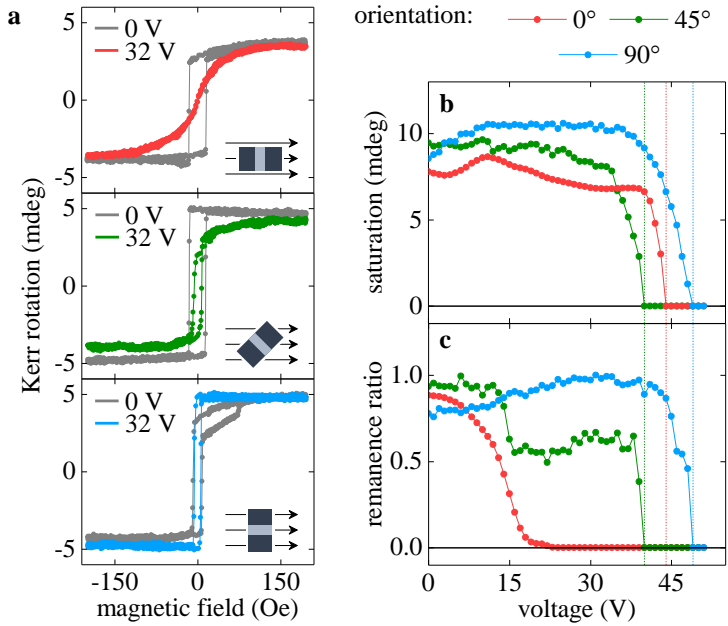

**Fig. 2 | Orientation dependence of voltage-controlled magnetism. a**, MOKE hysteresis loops recorded at equilibrium (0 V, grey lines) and during the resistive switching (32 V, color lines) using a magnetic field aligned along three different orientations with respect to the device length (schematically shown in pictograms). At zero voltage, the hysteresis loops have similar shape in all three orientations. As the applied voltage triggers the resistive switching, the hysteresis shape depends strongly on the orientation. **b, c**, MOKE signals at saturation (**b**) and remanence (**c**) as a function of the applied voltage for the three orientations. The voltage dependence of the saturation is nearly independent of the orientation. The complete suppression of the remanence at 0° vs. the remanence increase at 90° indicates the development of a strong uniaxial anisotropy during resistive switching. All measurements were performed at 100 K.

The resistive-switching-driven VCMA occurs without a considerable suppression of the FM state. We observed that the saturation signal in the MOKE loops, i.e., the peak-to-peak hysteresis loop amplitude, has a weak dependence on the applied voltage (Fig. 1d, orange line). Only when the applied voltage becomes large enough to drive the material into the PM state, the saturation signal plummets to zero. The MOKE signal at remanence, on the other hand, rapidly vanishes as the device undergoes resistive switching (Fig. 1d, green line). The fact that the remanence becomes zero without a noticeable change of the saturation suggests that the observed VCMA cannot be simply attributed to the applied voltage driving the device on the verge of the PM transition (for example, due to Joule heating) where the uniform magnetization state might not be stable. Instead, our measurements show evidence that, as the resistive switching progresses, an internal magnetic field develops along the device length. Fig. 1e compares the coercive field and the field required to achieve the saturation. At low voltages, the saturation field closely tracks the coercive field. At the onset of resistive switching at 8 V, the saturation field begins to grow rapidly and eventually reaches a maximum of ~100 Oe, which is the factor of ~5 increase compared to its initial zero-voltage value. Such a large increase of saturation field suggests that the external magnetic field has to overcome a strong opposing internal field. Because this internal field greatly exceeds the coercivity, the remnant uniform magnetization state along the device length becomes unstable leading to the observed development of the oblique, non-hysteretic MOKE loop.

MOKE measurements at different sample orientations provide an insight into the anisotropy change induced by the resistive switching in the LSMO device. Fig. 2a compares the hysteresis loops acquired with the magnetic field along 0°, 45° and 90° with respect to the device length. Similar to the measurements presented in the previous paragraphs, the loops were acquired at a spot close to one of the electrodes, i.e., away from the location where the PM insulating barrier initially forms during the resistive switching. At zero bias (grey loops in Fig. 2a), the hysteresis loops along all three orientations have square shape indicating that the device can be easily magnetized in any direction. This high loop squareness can be explained by weak magneto-crystalline anisotropy of LSMO[29] and the dominant contribution of the thin film shape anisotropy favoring the easy-plane anisotropy in equilibrium. During the resistive switching, the hysteresis loop appearance depends strongly on the device orientation with respect to the applied magnetic field (color loops in Fig. 2a). While the loops acquired at 0° and 45° show a similar trend – a suppression of the loop squareness and remanence under applied voltage, the loop acquired at 90°, on the contrary, displays the squareness and remanence enhancement. Figs. 2 b and c summarize the dependence of the MOKE signals at saturation and remanence on the applied voltage for the three orientations. The signal at saturation has similar voltage dependence in all three orientations (Fig. 2b). The saturation remains nearly unchanged until the critical voltage drives the device across the PM transition, at which points the saturation signal rapidly drops to zero. We note that because the sample was manually reattached and the MOKE optics had to be refocused for every orientation, the location of the MOKE measurements differed slightly, resulting in slight variation of the critical voltage. The orientation insensitivity of the voltage dependence of the MOKE signal at saturation can be explained by taking into account that the device temperature is controlled by Joule heating generated during the resistive switching that occurs in the same way irrespective of the orientation. The MOKE signal at remanence vs. voltage dependence, on the other hand, clearly shows the VCMA effect. Initially, the remanence signal is large for all three orientations. When the applied voltage triggers the resistive switching, the remanence to saturation ratio drops to zero for the 0° orientation, decreases to ~0.6-0.7 for 45° and becomes ~1 for 90°, suggesting a sinusoidal angular dependence. The orientation-dependent MOKE measurements, thus, indicate the development of a strong uniaxial anisotropy in the direction perpendicular to the device length (i.e. perpendicular to the current flow and parallel to the PM insulating barrier) when the applied voltage induces the metal-to-insulator switching in LSMO.

The VCMA effect is most efficient at low temperatures. Fig. 3a compares the hysteresis loops recorded at equilibrium ($V = 0$) and during the resistive switching ($V \neq 0$) at three temperatures. At 50 K, the applied voltage causes the complete suppression of the remanence and coercivity and strongly increases the saturation field leading to a highly oblique MOKE loop. At 200 K, a substantial decrease of the remanence and coercivity can be induced, however, the saturation field remains close to its initial value at zero voltage. At 300 K, the applied voltage has no noticeable impact on the hysteresis loop shape. The VCMA effect correlates well with the temperature behavior of the resistive switching (Fig. 3b). At low temperatures where the VCMA is strong, the I-V curves manifest a highly pronounced NDR region – a characteristic property of the resistive switching. As the temperatures increases, the NDR region first becomes shallow and then eventually disappears. This gradual loss of the strong nonlinear features in the I-V characteristics coincides with the decreasing effectiveness of the VCMA effect, i.e., the decreasing impact of voltage on the hysteresis loop shape with increasing temperature. The correlation between the electric and magnetic temperature behaviors provides strong evidence that the resistive switching is the driving force behind the observed VCMA effect in LSMO.

To quantitatively characterize the temperature dependence of the VCMA efficiency, we investigate the influence of the resistive switching on the remanence ratio (Fig. 3c) and saturation field (Fig. 3d). The remanence ratio at equilibrium is nearly independent of temperature and it is close to 1. By driving the device into the resistive switching, the remanence can be fully suppressed at temperatures below 150 K. At higher temperatures, only a partial remanence suppression can be achieved by the switching. Near room temperature, applying voltage causes no noticeable remanence change, which indicates that the change of the anisotropy is either small or absent. The maximum saturation field under applied voltage (Fig. 3d) has a similar trend as the remanence ratio, both parameters decrease with increasing temperature, however, there is an important distinction between them. The maximum saturation field has two distinct regimes. Below 125 K, resistive switching leads to a factor of 5 increase of the saturation field compared to its initial zero-voltage value. Above

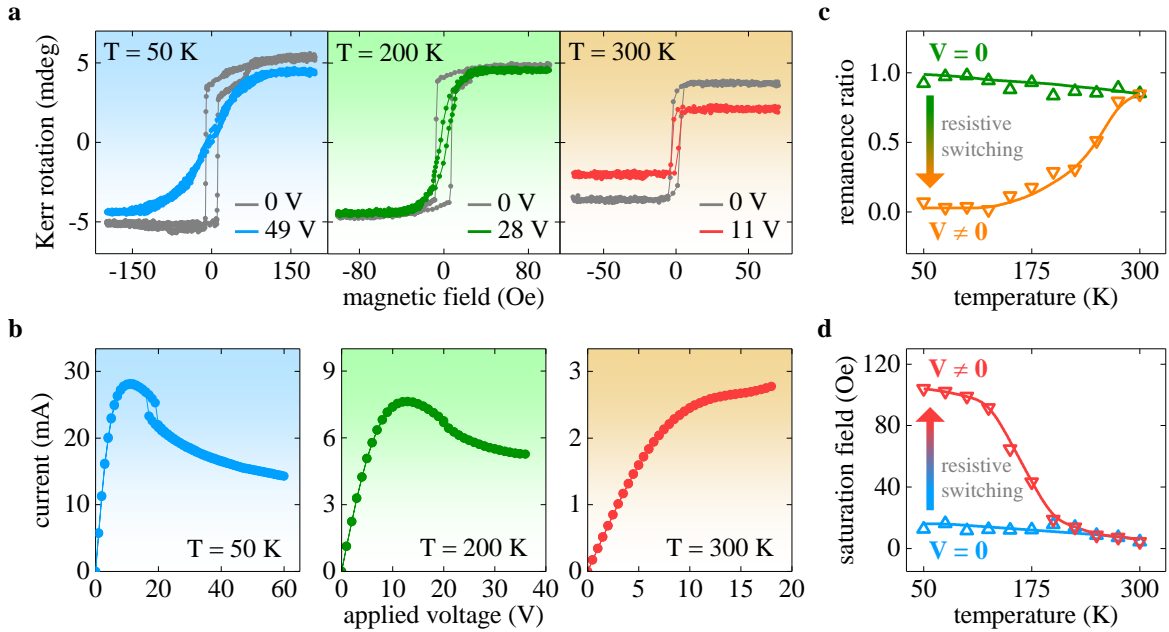

**Fig. 3 | Temperature dependence of the voltage-controlled magnetism. a**, MOKE hysteresis loops recorded at equilibrium (grey lines) and during the resistive switching (color lines) at three different temperatures. The VCMA effect during the resistive switching is strong at 50 K, moderate at 200 K, weak or absent at 300 K. **b**, I-V curves recorded at the same temperatures as the loops in **a**. As temperature increases, the nonlinear behavior associated with the resistive switching (especially the negative differential resistance) becomes less pronounced. **c**, Temperature dependence of the remanence ratio in equilibrium ($V = 0$, green curve) and during the resistive switching ($V \neq 0$, orange curve). At low temperatures, inducing the switching results in the suppression of the remanence ratio indicating strong VCMA effect. **d**, Temperature dependence of the saturation field in equilibrium ($V = 0$, blue curve) and during the resistive switching ($V \neq 0$, red curve). Strong VCMA effect at low temperatures is accompanied by a large increase of the saturation field. In **c** and **d**, the points are experimental values and lines are guides to the eye.

200 K, the saturation field remains nearly unchanged during the switching, even though a moderate VCMA effect persists to at least 275 K as observed in the remanence ratio vs. temperature dependence (Fig. 3c). This two-regime temperature behavior of the maximum saturation field could be an indication that multiple mechanisms contribute to the VCMA effect. We theoretically investigate one of the possible mechanisms using micromagnetic analysis.

Our micromagnetic simulations identified the important role of magnetostatic fields in the resistive switching-induced VCMA. We considered a 500×500×10 nm$^2$ FM slab subjected to a temperature gradient that emulates the effect of the PM insulating barrier formation inside the device during the switching (Fig. 4a). This model represent a FM half of the LSMO device that experiences strong heating from the PM barrier side, while the cold electrode provides cooling from the opposite side. Because the exact temperature profile in the experiment is unknown, we assumed a linear temperature gradient in our simulations in order to capture the basic physics. The temperature gradient produces a monotonic, but nonlinear decrease of the saturation magnetization along the device length (Fig. 4b). We used an experimental magnetization vs. temperature dependence of an unpatterned LSMO film to represent the magnetization distribution in the model. We achieved excellent qualitative agreement between the MOKE measurements and micromagnetic simulations. Without the temperature gradient, the simulated loop has a square shape (Fig. 4c, top) similar to the experimental loops recorded in equilibrium. With the temperature gradient, the simulated loop is oblique when the field is along the gradient (Fig. 4c, middle) and the loop has an increased squareness when the field is perpendicular to the gradient (Fig. 4c, bottom), in agreement with the orientation-dependent measurements in Fig. 2.

The simulations reveal that the high-temperature FM region (i.e., a region next to the hot PM barrier) controls the magnetization reversal of the entire slab. In this region, the saturation magnetization rapidly changes leading to the emergence of an effective magnetostatic charge density when there is a magnetization component along the temperature gradient, which, in turn, creates strong magnetostatic fields (Fig. 4d). These magnetostatic fields can reach ~100 Oe in the region next to the effective charge, exceeding the intrinsic coercivity. On the other hand, no magnetic charges or magnetostatic fields are generated when there is no magnetization component along the temperature gradient. The magnetization-orientation-dependent generation of magnetostatic charges acts as a shape anisotropy producing an easy axis perpendicular to the temperature gradient. Because of this anisotropy, the hysteresis loop squareness is high in the direction perpendicular to the temperature gradient (effective easy axis) and the squareness is low in the direction along the gradient (effective hard axis).

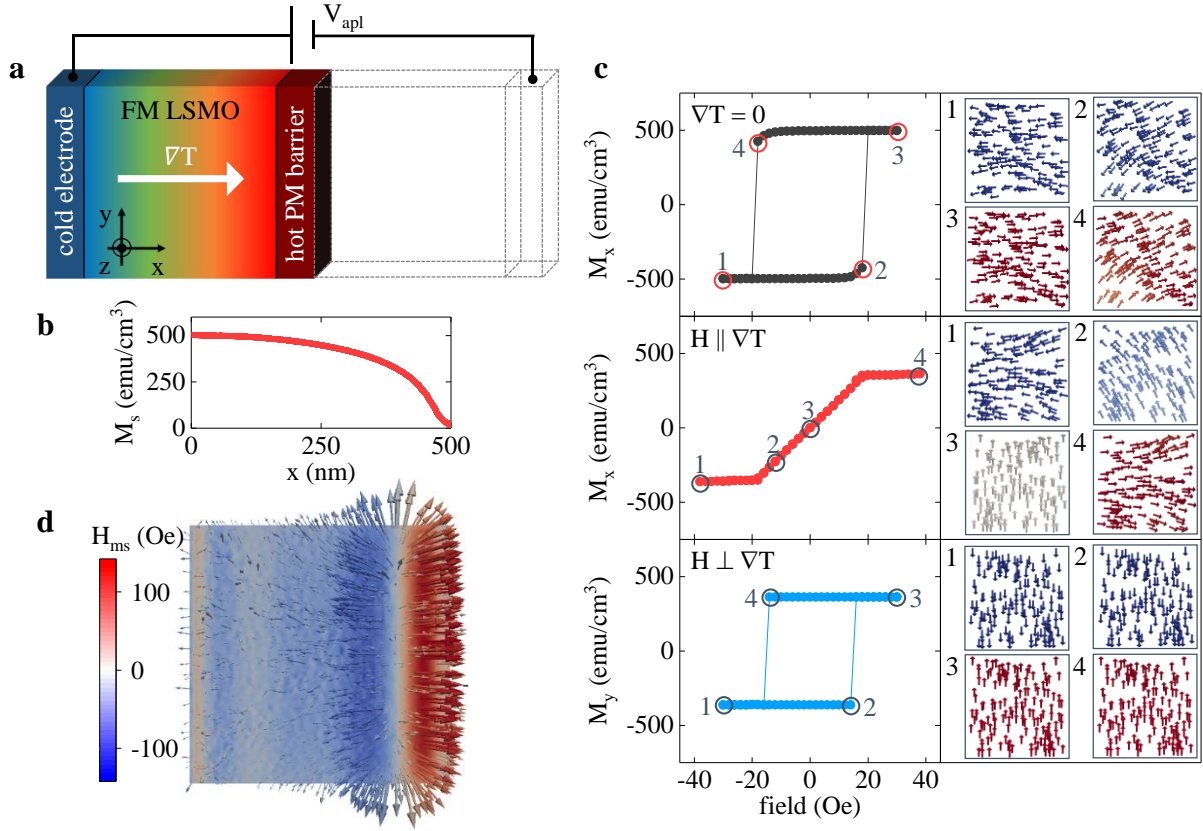

**Fig. 4 | Micromagnetic simulations of the voltage-controlled magnetism. a**, Model schematics: 500×500×20 nm$^2$ FM slab subjected to a linear temperature gradient. This temperature gradient emerges because the resistive switching in LSMO occurs by the formation of a PM barrier whose temperature is above $T_c$. **b**, Spatial distribution of the saturation magnetization used in the model to represent the effect of the temperature gradient. This distribution is based on the experimental magnetization vs. temperature dependence measured in an unpatterned LSMO film. **c**, The panels on the left show simulated hysteresis loops without the temperature gradient (top) and with the magnetic field parallel (middle) and perpendicular (bottom) to the temperature gradient. The change of the loop shape qualitatively reproduces the observed VCMA effect in the LSMO devices. The panels on the right show the calculated magnetization distribution at the numbered points in the hysteresis loops. **d**, Magnetostatic field distribution when the magnetization of the slab is saturated parallel to the temperature gradient (hot side on the right). Because rapid saturation magnetization change occurs in the hot temperature region, this region acts as a source of strong magnetostatic fields and it is responsible for the change of the hysteresis loop shape in **c**.

The results of our micromagnetic simulations are reminiscent of the behavior in magnetization-graded materials[30,31]. In such materials, a spatial variation of the chemical composition produces a spatial gradient of the magnetization, which has a strong impact on the magnetic anisotropy. In the case of resistive switching in a FM device, the magnetization grading emerges due to the natural tendency of resistive switching to occur in spatially inhomogeneous patterns, providing the ability to turn on/off and to adjust this grading at will using applied voltage.

The control of magnetic anisotropy by resistive switching demonstrated in this work potentially can be implemented in a large variety of magnetic systems. Volatile switching due to the triggering of a metal-insulator transition (similar to this work) can be induced in other members of the manganite family that have ferromagnetic and antiferromagnetic ordering[32,33], in antiferromagnetic $V_2O_3$[34] and rare-earth nickelates[35], etc. Nonvolatile switching due to the electrically induced ionic migration has been demonstrated in ferromagnetic $(La,Sr)CoO_3$[36], antiferromagnetic $SrFeO_3$[37], ferrimagnetic $Fe_3O_4$[38] and other magnetic oxides. Resistive switching, volatile or nonvolatile, often occurs by the formation of a spatially inhomogeneous pattern, such as a longitudinal conducting filament[26,27] or transverse insulating barrier[28,39], which locally "injects" a different electronic and magnetic phase into an otherwise homogeneous material. Similar to the presented results for LSMO, the formation of a filament/barrier can enable voltage-controlled magnetism in other resistive switching materials. The interaction between resistive switching and magnetic order could be especially important for antiferromagnetic materials in which an efficient way to control Néel vector is the key ingredient to enable a new class of spintronic devices[40].

## Acknowledgements

The magnetism research was supported by the U.S. Department of Energy (DOE), Office of Science, Basic Energy Sciences (BES), Materials Sciences and Engineering Division under Award # DE-FG02-87ER-45332. Material synthesis and numerical simulation were supported as a part of Quantum Materials for Energy Efficient Neuromorphic Computing (Q-MEEN-C), an Energy Frontier Research Center funded by the DOE, Office of Science, BES under Award # DE-SC0019273.

## Methods

**Sample preparation**. 20 nm thick $La_{0.7}Sr_{0.3}MnO_3$ films were grown epitaxially on a (001)-oriented $SrTiO_3$ substrates using pulsed laser deposition at the laser fluence of 0.7 J/cm$^2$ and frequency of 1 Hz. The substrate temperature was 700 °C and the oxygen pressure was 0.3 Torr during the growth. After the growth, the films were slowly cooled to room temperature in 300 Torr $O_2$ atmosphere. The devices were made by fabricating (100 nm Au)/(20 nm Pd) electrodes using photolithography and e-beam evaporation and patterning the film in Ar plasma. The 50×100 μm$^2$ devices we patterned along [100] crystallographic direction.

**Measurements**. Magnetic measurements were performed using a Montana Instruments NanoMOKE 3 system equipped with a 660 nm laser. The laser spot was ~5 μm. The magnetic field was cycled at 2.3 Hz repetition rate. The voltage was applied *in-operando* using a Keithley 2450 source meter. For the angular-dependent measurements (Fig. 2), the sample was manually aligned and re-attached to the sample holder for each orientation. Magnetization vs. temperature dependence of an unpatterned film was measured using a vibrating sample magnetometer in a Quantum Design PPMS Dynacool system.

**Simulations**. The micromagnetic simulations of a 500×500×10 nm$^2$ ferromagnetic slab were performed in FastMag. The exchange constant was 10 pJ/m, while all intrinsic anisotropies were set to zero. To model a large film, the surface magnetic charges on the sides of the slab were removed from the calculations of magnetostatic fields. The magnetic surface charges on the top and bottom of the slab produced the magnetization in-plane state. A small out-of-plane roughness of 0.5 nm was introduced to maintain the magnetization at remanence and to obtain reasonable values of coercive field and loop squareness in the absence of internal anisotropies. The effect of the temperature gradient was emulated by introducing the saturation magnetization variation along the length of the slab using the experimental magnetization vs. temperature data recorded for an unpatterned LSMO film.